\def\gapprox{\lower.4ex\hbox{$\;\buildrel >\over{\scriptstyle\sim}\;$}}
\def\lapprox{\lower.4ex\hbox{$\;\buildrel <\over{\scriptstyle\sim}\;$}}

  \def\selectedoptions{final}

\documentclass[
   \selectedoptions
  ]
  {aipproc}

\layoutstyle{6x9}

\SetInternalRegister\hbadness{8000} 

\begin{document}

\title 
      [Highest-energy cosmic ray acceleration]
      {On a mechanism of highest-energy cosmic ray 
      acceleration}


\author{C. Litwin}{
  address={Department of Astronomy \& Astrophysics, The University of Chicago, 5640 S Ellis Avenue, Chicago, IL 60637},
  email={litwin@zohar.uchicago.edu}
}

\iftrue
\author{R. Rosner}{
  address={Department of Astronomy \& Astrophysics, The University of Chicago, 5640 S Ellis Avenue, Chicago, IL 60637},
  email={rrosner@oddjob.uchicago.edu}}

\fi

\copyrightyear  {2001}

\begin{abstract}
A recently proposed mechanism of acceleration of highest energy cosmic 
rays by polarization electric fields arising in plasmoids injected 
into neutron star magnetospheres is discussed.
\end{abstract}

\date{\today}

\maketitle

\section{Introduction}\label{overview}

The problem of the origin of ultra-high-energy cosmic rays (UHECR) -
those with energy $\gapprox 10^{19}$ eV - continues to pose a serious
theoretical challenge (Hillas 1984, Biermann 1997, Cronin 1999,
Bhattacharjee \& Sigl 2000, Olinto 2000).  No convincing explanation
that could account for all main observables -- energy, spectrum and
flux -- has been offered to-date.  Of particular interest are cosmic
rays in the highest energy range, above $\gapprox 5\times 10^{19}$ eV.
This radiation does not exhibit any significant anisotropy connected
with the galactic disk and is therefore generally assumed to be of
extragalactic origin (Blandford 2000).  Moreover, the seeming, albeit
uncertain, change of slope of the energy spectrum at $\sim 5\times
10^{19}$ eV is frequently taken as an indication of the appearance of
a new, distinctly different (and presumably extragalactic) component
of the spectrum.  At the same time the UHECR do not exhibit any sign
of Greisen-Zatsepin-Kuzmin (GZK) cut-off (Greisen 1966, Zatsepin \&
Kuzmin 1966).  Thus if carried by singly charged particles, light
nuclei or photons, this radiation would need to originate at distances
$\lapprox 50$ Mpc.  Nevertheless, the direction of the incoming
radiation does not appear to be correlated with any plausible sources
within this distance.

Existing theories of the UHECR generation are usually put into two
general categories: the ``bottom-up'' scenarios, in which particles
are accelerated from lower energies to the ultra-high energies; and
the ``top-down'' scenarios in which particles are ``born'' with
ultra-high energies in a decay of some ultra-massive X particles,
usually relics of the early universe.

Top-down scenarios, in addition to relying on uncertain physics, face
difficulties explaining both the flux and the energy spectrum of UHECR
(see Olinto 2000).  The primary difficulty for the bottom-up
scenarios is the acceleration mechanism.

Acceleration scenarios are generally divided into two types (cf. 
Hillas 1984): (1) direct acceleration, by electric fields; or (2)
statistical Fermi acceleration by shocks in magnetized plasmas.

Statistical Fermi acceleration by supernova shocks (Axford et al. 
1977, Krymsky 1977, Bell 1978, Blandford \& Ostriker 1978) is
considered a source of the cosmic rays below the ``knee'' ($\sim
5\times 10^{15}$ eV), which are believed to be of galactic origin.  It
gives rise to a power law energy spectrum, which combined with the
energy dependence of the diffusion coefficient, as inferred from the
data on secondary nuclei, yields the energy spectrum similar to the
observed one.  This mode of acceleration becomes inefficient at higher
energies (Lagage \& Cesarsky 1983).  

The primary difficulty with the direct acceleration scenarios is the
existence of sufficiently large voltages.  Most commonly considered
sources are unipolar inductors, such as rapidly spinning magnetized
neutron stars or blackholes.  In the case of pulsars, the rotation
gives rise an emf too small to accelerate iron nuclei to the UHECR
energies (Berezinskii et al.  1990, Blandford 2000).  A spinning
blackhole in the center of a radiogalaxy generates an emf sufficient
to accelerate protons to energies $10^{19}-10^{20}$ eV. A difficulty
with this scenario, however, is the presence of a dense pair plasma
and intense radiation which would cause energy losses of accelerated
particles.  Another argument frequently used (e.g., Hillas 1984)
against direct acceleration scenarios is that it is not clear how the
power-law energy spectrum, characteristic for cosmic rays, could emerge.

\section{Acceleration mechanism}

We have recently proposed (Litwin \& Rosner 2001) an alternative
scenario of a galactic or galactic-halo origin of the UHECR. This
scenario goes some way toward overcoming some of the main difficulties
described in the previous section.  We describe this recent work in
the present section.

We started with the observation that polarization electric fields
arise in plasma ``blobs'' or streams injected at large angles into the
magnetic field (Chandrasekhar 1960, Schmidt 1960; more recently,
Litwin, Rosner \& Lamb 1999 and refs.  therein).  The appearance of
this electrostatic field allows for the plasma stream to penetrate
into the magnetic field, as has been demonstrated in numerous
laboratory experiments (Baker \& Hammel 1965, Meade 1965, Lindberg
1978) and numerical simulations (Galvez \& Borovsky 1991, Neubert et
al.  1992).

Outside the plasma, the electrostatic field has an approximately
dipolar character and has a nonvanishing component along the magnetic
field.  This field accelerates particles (electrons and ions) out of
the plasma and gives rise to the plasma current that leads to the
eventual halting of the plasma cross-field motion.  The energy of
accelerated particles is $\sim ZeU$ where $Ze$ is the particle charge
and $U$ is one-half of the electrostatic potential difference across
the plasma stream.  This phenomenon of particle acceleration and the
above estimate of particle energy has been verified in experiments
(Lindberg 1978) and in computer simulations (Galvez \& Borovsky 1991,
Neubert et al.  1992).

For a sufficiently large stream width (much greater than the Larmor
radius corresponding to the stream velocity), the energy of an
accelerated particle can greatly exceed the kinetic energy of a bulk
plasma particle.  In particular, for a plasmoid of width $\sim 10$ km,
free-falling onto a neutron star with the surface magnetic field $B\sim
10^{13}$ G, the voltage $U\sim 10^{21}$ V.

We applied the above-described basic physics to a plasmoid injected
into a neutron star magnetosphere.  Such plasmoids may results from
planetoid impacts onto neutron stars of the type that has been
previously discussed in the literature as possible sources of galactic
gamma ray bursts (Colgate \& Petschek 1981, Lin et al.  1991, Katz et
al.  1994, Wasserman \& Salpeter 1994, Colgate \& Leonard 1996).  The
motion at distances greater than the Alfv\'en radius (i.e., where the
ram pressure is equal the magnetic pressure) is unaffected by the
magnetic field.  Following Colgate \& Petschek (1981) to describe the
process of break-up, and subsequent compression and elongation, of an
iron planetoid by tidal forces the density and the size of the
impacting plasmoid at the Alfv\'en point is determined.  During the
infall the planetoid matter becomes ionized by the motional electric
field.  In the vicinity of the Alfv\'en radius it is assumed, as is
customary (e.g., Lamb et al.  1973), that the magnetic field
penetrates the plasma (due to, e.g., anomalous resistivity).  The
subsequent cross-field motion leads to plasma polarization, as
described by Chandrasekhar (1960) and Schmidt (1960).  The magnitude
of the accelerating potential due to this polarization electric field
is then determined by the energy of particles, accelerated along the
magnetic field by the polarization field, from the plasmoid velocity,
its size and the magnetic field at the Alfv\'en radius.

Subsequently we determined the energy of iron nuclei emerging from the
magnetosphere, by solving numerically the particle equation of motion
in the dipole magnetic field taking into account the radiation
reaction force.  As expected, the energy of emerging particles is much
smaller than the initial energy unless they are accelerated close to
the magnetic axis.

The energy of emerging particles is quite sensitive to the angle
between the particle trajectory and the magnetic axis.  From this, the
energy spectrum of cosmic rays generated in a single impact event can
be deduced (Litwin \& Rosner 2001).  For the range of parameters
considered (magnetic field $B\sim 10^{12}-10^{14}$ G, planetoid mass
$M_{p}\sim 10^{22}-10^{24}$ g), the spectrum depends weakly on the
magnetic field and the planetoid mass.  To a good approximation the
emerging particles have a power-law energy spectrum: ${dN}/{dE}\sim
E^{\mu}$.  The value of the exponent $\mu\sim 2.9-3.0$ for magnetic
fields in the considered range is within the measurement uncertainty
of the value found by AGASA (Takeda et al.  1998).

The number of particles at given energy produced in a \emph{single}
event can be found (Litwin \& Rosner 2001) from the energy spectrum
and from the total charge carried by accelerated particles.  The
latter is determined by integrating the equation of motion.  For an
iron planetoid, with the fiducial mass of $10^{23}$ g, the number of
particles with energies exceeding $10^{19}$ eV is $\sim 2-14\times 10^{28}$.

\section{Magnetic containment, energy spectrum and flux of
UHECR}\label{confinement}

Larmor radii of iron nuclei with energies less than $\lapprox 10^{20}$
eV in a magnetic field with strength 3-10 $\mu$G, characteristic of
the galactic magnetic field (Kronberg 1994), are less than $\sim 1$
kpc.  Thus iron nuclei constituting UHECR would be confined by the
galactic magnetic field with a characteristic gradient length scale of
10 kpc.  Assuming a steady state, the energy spectrum observed on
Earth differs from the source spectrum if the confinement time is a
function of energy.  It is usually believed (see, e.g., Sigl \&
Bhattacharjee 2000) that confinement time decreases with increasing
energy.  Indeed, the chemical composition of CR in the $1-10^{3}$ GeV
per nucleon range can be interpreted as a power law behavior of the
diffusion coefficient $D(E)\sim E^{\mu}$, with $\mu\sim 0.3-0.7$ (see
Berezinskii et al.  1990).  A power law behavior of the diffusion
coefficient results also in theoretical models, such as the model of
Jokipii (1975) in which particles are scattered by the magnetic field
fluctuations which vary only in the direction transverse to the mean
field.  If the particle Larmor radius is much smaller than the
integral scale $L_{c}$, a power-law fluctuation spectrum, with
exponent $\alpha$, results in a power-law dependence of the diffusion
coefficient on energy, with the exponent $\mu =(1-\alpha )/2$.  In
particular, for the Kolmogorov spectrum ($\alpha =-5/3$), $\mu =4/3$.

However, the UHECR confinement time dependence on energy may be
qualitatively different.  First, the small Larmor radius approximation
may be inapplicable to the highest energy cosmic rays.  If one assumes
that the integral scale of the galactic turbulence is 100 pc (Parker
1979), and that the galactic magnetic field is in the range 3-10
$\mu$G (Kronberg 1994), the Larmor radius of iron nuclei is larger
than the integral scale for energies higher than $1-3\times 10^{19}$
eV. In this regime, the dependence of the cross-field transport on
energy might be significantly different.  As a specific example, the
previously discussed model of Jokipii (1975) yields in this regime
$\mu =-1/2$.  Thus the confinement times would increase with energy,
assuming that the latter were determined by the cross-field diffusion. 
On the other hand, if the confinement time were determined by the
motion along the magnetic field, it would be independent of energy on
the galactic/galactic halo length scales for the ultrarelativistic,
collisionless particles in the UHECR energy range.

If the confinement time is known, the rate of planetoid impact events
required to generate the observed UHECR flux on Earth can be
determined.  The upper bound on the confinement time is given by the
rate of photodisintegration on the infrared radiation background
(Stecker 1998).  The lower bound can be obtained from the escape time
from the galactic magnetic field at the velocity of the curvature
drift, assuming a 10 kpc curvature radius.  For $10^{19}$ eV iron
nuclei, this escape time is $\sim 10^{13}$ s; the escape time is
longer if the field is axisymmetric and possesses closed flux surfaces. 
A similar estimate is found for the transit time along a spiral
magnetic field twisting by 4$\pi$ within radial distance of 10 kpc. 
Thus it is reasonable to expect that the confinement time will be in
the range $\sim 10^{13}-10^{16}$ s.  Then assuming that the density of
neutron stars is $2\times 10^{{-3}}$ pc$^{{-3}}$ (Shapiro \& Teukolsky
1983) the observed flux of UHECR results if the impact rate is $\sim
10^{-4}-10^{-8}$ yr$^{-1}$ on each neutron star.

One can speculate whether such an impact rate is plausible.  If one
assumes (Lin et al. 1991, Nakamura \& Piran 1991, Colgate \& Leonard
1996) that the planetoids originate in the accretion disk, formed from
the estimated $10^{29}-10^{32}$ g of matter captured by the neutron
star following the supernova explosion, then the rate of accretion of
$10^{23}$ g planetoids in a galaxy cannot exceed $\sim 10^{4}-10^{7}$
yr$^{-1}$, assuming the neutron star birth rate to be $10^{-2}$
yr$^{-1}$.  Since $\sim 10^{29}$ particles with energies higher than
$10^{19}$ eV are generated in each impact event, the upper bound on
the rate of UHECR generation is $\sim 10^{33}-10^{36}$ yr$^{-1}$. 
From the observed flux (Takeda et al. 1998) it follows that the
density of particles in this energy range is $6\times 10^{-29}$
cm$^{-3}$.  Assuming that the radiation is confined in a sphere of 10
kpc radius, this would require a confinement time exceeding
$10^{4}-10^{7}$ years.  The estimated confinement time, mentioned in
the previous paragraph, is greater than or comparable to this lower
bound.

\section{Conclusions}

In this talk we reviewed a model (Litwin \& Rosner 2001), which can
potentially solve some of the problems, discussed in the Introduction. 
This model is an example of a direct acceleration process that
explicitly results in a power-law energy spectrum of UHECR. The source
spectrum agrees, within experimental uncertainties, with the observed
spectrum.  While the latter has not been calculated, it is plausible
that the energy dependence of the confinement time will not result in
a steepening of the spectrum in the UHECR range of energies.  Also the
magnitude of the observed UHECR flux results from a plausible rate of
impact events; and the resulting spectrum may be only weakly
anisotropic both due to the confining effect of the magnetic field and
because the sources are neutron stars both in the galactic disk and in
the galactic halo (cf.  Bulik, Lamb \& Coppi 1998).

\begin{theacknowledgments} The authors thank Attilio Ferrari, Roger
Hildebrand, Don Lamb, Angela Olinto and Simon Swordy for illuminating
discussions. Useful comments by Pasquale Blasi, Willy Benz, Sterling
Colgate, Walter Drugan, Carlo Graziani, Cole Miller, Don Rej and Eli
Waxman are also gratefully acknowledged.  This research has been
supported by the Center for Astrophysical Thermonuclear Flashes at the
University of Chicago under Department of Energy contract B341495.
\end{theacknowledgments}

\bibliographystyle{aipproc}

\end{document}